\documentclass[conference,10pt]{IEEEtran}

\usepackage{algorithmic,amsmath,amssymb,amsthm,array,bbm,bibentry,cite,color,comment,dsfont}
\usepackage{enumerate,eurosym,float,graphicx,lettrine,mathrsfs,multirow,nomencl}
\usepackage{pict2e,psfrag,ragged2e,relsize,setspace,subfigure,supertabular,tabularx,url}
\usepackage[english]{babel}

\DeclareGraphicsExtensions{.eps}

{		\newtheorem{theorem}{Theorem}}
{ 		}
{ 		\newtheorem{lemma}{Lemma}}
{ 		}
{ 		\newtheorem{remark}{Remark}}
{		}
{		}
{ 		}

\renewcommand{\d}{\mathrm{d}}

\newcommand{\E}{\mathbf{E}}

\renewcommand{\P}{\mathbf{P}}

\newcommand{\setA}{\mathcal{A}}

\newcommand{\setK}{\mathcal{K}}

\newcommand{\setM}{\mathcal{M}}

\newcommand{\setS}{\mathcal{S}}

\newcommand{\as}{\textnormal{\tiny{as}}}

\newcommand{\nop}[1]{}

\def\P{\mathbb{P}}
\def\E{\mathbb{E}}

\begin{document}

\title{Delay Performance of Multi-Antenna Multicasting in Wireless Networks}

%
%

\author{\IEEEauthorblockN{Marios~Kountouris and Apostolos~Avranas} %
	
	\IEEEauthorblockA{Mathematical and Algorithmic Sciences Lab\\ Paris Research Center, Huawei Technologies Co., Ltd.\\20 Quai du Point du Jour, 92100 Boulogne-Billancourt, France} %
	Emails: \{apostolos.avranas,marios.kountouris\}@huawei.com
}

\maketitle

\begin{abstract}
Low-latency communication is currently attracting significant attention due to the emergence of mission-critical Internet of Things (IoT) applications and content-centric services. 
A deep understanding of the delay performance is essential for efficient wireless system design and end-to-end latency guarantees. 
In this paper, we investigate the network-layer performance of physical layer multi-antenna multicasting, i.e., when the same data is simultaneously conveyed to multiple users. 
We provide a statistical characterization of the service process in terms of its Mellin transform and derive probabilistic delay bounds using tools from stochastic network calculus. Furthermore, using extreme value theory, we characterize the service process for very large number of users and derive scaling laws as the number of antennas and/or users is taken to infinity. Our results can be used for system dimensioning to guarantee the delay requirements in wireless multicast networks.
\end{abstract}

\vspace{1mm}


\section{Introduction} \label{sec:intro}
The unprecedented data traffic growth over the last decade has radically transformed the wireless ecosystem. Two major trends related to traffic consumption could be identified. First, the largest amount of data traffic over the network requires high bandwidth and contains rich multimedia services, including video/audio streaming, cell broadcasting, and mobile television. Second, the same digital content is often requested simultaneously by or is of interest to a group of users, e.g., broadcasting of sporting events, popular videos, live shows, headline news, satellite broadcast, etc.
Several standards, such as 3GPP eMBMS (evolved Multimedia Broadcast Multicast Services) \cite{eMBMS} and DVB-H (Digital Video Broadcasting - Handheld) \cite{DVB}, have been introduced as a means to support efficient massive content delivery and multicast applications. Among the various transmission techniques that serve those objectives, physical layer multicasting (PLM) stands as a key enabler. The simplest scenario of PLM consists of a transmitter conveying a common message to a group of receivers, while more complex scenarios involve simultaneous transmissions of distinct messages to multiple multicast groups. 

Fifth generation (5G), the next generation mobile communication system, aims to support a broader spectrum of use cases than just mobile broadband. 5G envisions to provide wireless connectivity for massive machine-type communications (mMTC) and to support ultra-reliable, low latency communication (URLLC) for mission-critical services. Physical layer multicasting is envisaged to play a significant role in providing quality of service (QoS) in emerging 5G networks, especially with the anticipated integration of satellite communications in 5G terrestrial networks. Many mission-critical Internet of Things (IoT) applications and content-centric services can benefit from multicasting and its content diversity capabilities. PLM can also be used in edge caching, bringing content closer to the user in order to achieve the 5G low latency requirement. Prior work on PLM has mainly focused on its capacity limits \cite{Jindal,Park} and on beamforming techniques \cite{SidiropoulosPMC,KaripidisMC,CE_PLM}.

In this paper, we investigate the delay performance of physical layer multicasting in multiuser multiple-input single-output (MISO) downlink channels. We consider a low-complexity technique that does not require channel state information (CSI) at the transmitter and transmits using a spatially white covariance. We provide a statistical characterization of the service process in terms of its Mellin transform and derive bounds on the delay violation probability using tools from stochastic network calculus \cite{Chang00,Multihop13_Infocom,Fidler15_Guide}. Furthermore, using extreme value theory, we characterize the service process for increasing number of users and provide scaling laws as the number of antennas and/or users is taken to infinity. The analytical expressions based on the exact and the asymptotic distribution of the instantaneous channel gain quantify the effect of transmit power, number of transmit antennas and users on the delay distribution of physical layer multicasting. 

\section{System Model} \label{sec:syst}
We consider multicast data transmissions, i.e., a point-to-multipoint communication channel where the base station (BS) broadcasts common messages to all active users. The BS is equipped with $M$ antennas and serves $K$ single-antenna users. 

\subsection{Signal model}
We consider a flat-fading channel and assume that time is divided into equally sized time slots. The discrete-time complex baseband signal received by user $k$ at slot $i$ is given by
\begin{eqnarray}
y_{k,i} = \mathbf{h}_{k,i}\mathbf{x}_{i} + z_{k,i}, \ \ k=1, \ldots, K
\end{eqnarray}
where $\mathbf{h}_{k,i} \in \mathbb{C}^{1\times M}$ is the channel between the transmitter and $k$-th user at slot $i$, $\mathbf{x}_i \in \mathbb{C}^{M\times1}$ is the transmitted signal with $\mathbb{E}[\mathbf{x}^H\mathbf{x}] \leq 1$, and $z_{k,i}$ is zero-mean circularly symmetric complex Gaussian additive noise with variance of $1/P$.
We assume a Rayleigh block fading model, thus $\mathbf{h}_{k,i} \sim \mathcal{CN}(\mathbf{0},\mathbf{I}_M)$. 

We focus on low-complexity transmission techniques with no CSI at the transmitter and perfect CSI at the receiver. For that, a spatially white transmit covariance $\mathbf{Q}_i \triangleq \mathbb{E}[\mathbf{x}_i\mathbf{x}_i^H] = \frac{1}{M}\mathbf{I}_M$ is employed, fixed over all channel realizations and slots. Therefore, the instantaneous signal-to-noise ratio (SNR) for user $k$ in the $i$-th slot is given by $\gamma_{k,i} = \rho\|\mathbf{h}_{k,i}\|^2$, where $\rho = P/M$.

\subsection{Traffic Model}
The analysis follows a system-theoretic stochastic network calculus approach as in \cite{Multihop13_Infocom}, which involves a queueing system with stochastic arrival and departure processes described by bivariate stochastic processes $A(\tau, t)$ and $D(\tau, t)$, respectively. 
We consider a fluid-flow traffic model and the system starts with empty queues at $t=0$. 

The cumulative arrival and departure processes for any $0 \leq \tau \leq t$ during time interval $[\tau,t)$ are defined respectively as
\begin{eqnarray}
	A(\tau, t) = \displaystyle \sum_{i = \tau}^{t-1}a_i, \ \ D(\tau, t) = \displaystyle \sum_{i = \tau}^{t-1}d_i
\end{eqnarray} 
where $a_i$ models the number of bits that arrives at the queue at time instant $i$ and $d_i$ describes the number of bits that arrives successfully at the destination. For a successful transmission, the service process $C_i$ should be less or equal to the instantaneous achievable rate. In case of transmission errors, the service is considered to be zero as no data is removed from the queue. 

For lossless first-in first-out (FIFO) queueing systems, the delay $\Delta(t)$ at time $t$, i.e., the number of slots it takes for an information bit arriving at time $t$ to be received at the destination, is defined as 
\begin{eqnarray}
\Delta(t) = \inf\{u \geq 0 : A(0,t)/D(0,t+u) \leq 1 \}.
\end{eqnarray} 
The delay violation probability is given by 
\begin{eqnarray}
\Lambda(w,t) = \displaystyle \sup_{t\geq 0}\mathbb{P}\left[\Delta(t) > w \right].
\end{eqnarray}

\subsection{Service Process}
Assuming Gaussian codebooks and ideal link adaptation, the instantaneous transmission rate $C_i$ at time instant $i$ is equal to $C_i = N\log(1+\gamma_i)$ nats/s, where $N$ is the number of transmitted symbols per time slot (bandwidth) and $\gamma_i$ is the instantaneous SNR using multicasting. For exposition convenience, the rate is expressed with the natural logarithm.

The service process (or cumulative capacity) through period $[\tau,t)$ is defined as 
\begin{eqnarray}
S(\tau, t) \triangleq \displaystyle \sum_{i = \tau}^{t-1}C_i = \sum_{i = \tau}^{t-1}N\log(1+\gamma_i).
\end{eqnarray} 

\section{Delay Performance: Exact Analysis} \label{sec:perf_exact}
In this section, we provide a statistical characterization of the arrival and service processes in terms of their Mellin transforms as a means to obtain bounds on the delay violation probability. For fading channels, it is more convenient to map and analyze these processes into a transfer domain, referred to as exponential or SNR domain \cite{Multihop13_Infocom}.

First, we convert the cumulative processes from the bit domain to the SNR domain through the exponential function. The corresponding processes, denoted by calligraphic letters, are
\begin{equation*}
\mathcal{A}(\tau, t) = e^{A(\tau, t)}, \quad \mathcal{D}(\tau, t) = e^{D(\tau, t)}, \quad \mathcal{S}(\tau, t) = e^{S(\tau, t)}.
\end{equation*}

An upper bound on the delay violation probability can be computed as \cite{Multihop13_Infocom}
\begin{equation}\label{eq:delay_bound_1}
p_\mathrm{v}(w)  = \inf_{s>0}\left\lbrace K(s,-w)\right\rbrace \geq \Lambda(w)
\end{equation}
where $K(s,-w)$ is the so-called steady-state kernel, defined as
\begin{equation}\label{eq:ker_limits}
\mathcal{K}(s,-w) = \lim_{t\to\infty} \sum_{u=0}^{t}\mathcal{M}_{\mathcal{A}}(1+s,u,t)\mathcal{M}_{\mathcal{S}}(1-s,u,t+w)
\end{equation}
where $\mathcal{M}_{\mathcal{X}}(s,\tau,t) = \mathcal{M}_{\mathcal{X}_{(\tau,t)}}(s) = \E\left[\mathcal{X}^{s-1}(\tau,t)\right]$ denotes the Mellin transform of a nonnegative random variable for any $s \in \mathbb{C}$ for which the expectation exists.

\subsection{Mellin transform of arrival and service processes} 
Assuming that $\mathcal{A}(\tau, t)$ has stationary and independent increments, the Mellin transform becomes independent of the time instance, as follows
\begin{eqnarray}
\setM_\setA (s,\tau,t) & = & \E\left[\left(\prod_{i=\tau}^{t-1} e^{a_i}\right)^{s-1}\right] \nonumber \\ & = & \E\left[e^{a(s-1)}\right]^{t-\tau} = \setM_\alpha(s)^{t-\tau}
\end{eqnarray}
where we have defined $\alpha = e^a$. 
We consider the traffic class of $(z(s), \lambda(s))$-bounded arrivals, whose moment generating function in the bit domain is bounded by \cite{Chang00}
\begin{eqnarray}
\frac{1}{s}\log\E[e^{sA(\tau,t)}] \leq \lambda(s)\cdot(t-\tau) + z(s) 
\end{eqnarray}
for some $s>0$. Here we consider the case where $\lambda$ is independent of $s$ and $z(s) = 0$, thus we have  
\begin{equation}\label{eq:defmellin_alpha}
\setM_\alpha(s) = e^{\lambda(s-1)}.
\end{equation}

Since $C_i$ is i.i.d., the Mellin transform of the cumulative service process with $g(\gamma_i) = 1+\gamma_i$ is 
\begin{eqnarray}\label{eq:defmellin_s}
\setM_\setS(s,\tau,t) &=& \E\left[\left(\prod_{i=\tau}^{t-1}g(\gamma_i)^N\right)^{s-1}\right] \nonumber \\ &=&  \E\left[g(\gamma)^{N(s-1)}\right]^{t-\tau}  \nonumber \\
&=& \setM_{g(\gamma)}\left(1+N(s-1)\right)^{t-\tau}.
\end{eqnarray}

\subsection{Delay Bound}\label{sec:delay_viol}
Plugging (\ref{eq:defmellin_alpha}) and (\ref{eq:defmellin_s}) into (\ref{eq:ker_limits}) and following \cite{Multihop13_Infocom}, the steady-state kernel can be finally rewritten as 
\begin{eqnarray}
\mathcal{K}(s,-w) = \frac{\left(\mathcal{M}_{g(\gamma)}(1-Ns)\right)^{w}}{1 - \mathcal{M}_{\alpha}(1+s)\mathcal{M}_{g(\gamma)}(1-Ns)},
\end{eqnarray} 
for any $s > 0$ under the stability condition $\mathcal{M}_{\alpha}(1+s)\mathcal{M}_{g(\gamma)}(1-Ns) < 1$. The delay bound (\ref{eq:delay_bound_1}) thus reduces to
\begin{equation}
p_\mathrm{v}(w) = \inf_{s>0}\left\lbrace\frac{\left(\mathcal{M}_{g(\gamma)}(1-Ns)\right)^{w}}{1 - \mathcal{M}_{\alpha}(1+s)\mathcal{M}_{g(\gamma)}(1-Ns)}\right\rbrace.
\end{equation}

\subsection{Service for Physical Layer Multicasting}
In this section, we derive exact closed-form expressions for the steady-state kernel $\setK(s,-w)$ of multi-antenna multicasting. For that, we need to derive the Mellin transform of $g(\gamma)$, which is a function of the instantaneous SNR. For exposition convenience, we set $N=1$ and we drop the time subindex since SNRs are independent and ergodic. 

Since the common message should be decoded by all $K$ users, the instantaneous rate should not exceed the rate achievable by the weakest user. Therefore, the instantaneous SNR of the system is given by $\gamma_i = \rho\displaystyle \min_{1\leq k \leq K}\|\mathbf{h}_k\|^2$, where $X_k \triangleq \|\mathbf{h}_k\|^2 \sim \chi_{2M}^2$ follows a chi-squared distribution with $2M$ degrees of freedom. 
The CDF of $X_{(1)} \triangleq \displaystyle \min_{1\leq k \leq K}X_k$ is 
\begin{eqnarray}
\label{CDFmin}
F_{X_{(1)}}(x) = 1 - (1-F_{X}(x))^K = 1 - \left(\frac{\Gamma(M,x)}{\Gamma(M)}\right)^K  
\end{eqnarray}
where $\Gamma(a,x) = \int_{x}^{\infty}t^{a-1}e^{-t}\,\mathrm{d}t$ and $\Gamma(a) = \Gamma(a,0)$ is the upper incomplete and complete gamma function, respectively.
\smallskip

\noindent The Mellin transform of $g(\gamma)$ is given by
\begin{eqnarray}
\mathcal{M}_{g(\gamma)}(s) = \E\left[g(\gamma)^{s-1}\right] = \displaystyle \int_{0}^{\infty}(1+\rho x)^{s-1}\d F_{X_{(1)}}(x).
\end{eqnarray}

Using (\ref{CDFmin}) and the multinomial theorem, and after some algebraic manipulations, we obtain the following result.

\begin{theorem}\label{Th1:main}
	For physical layer MISO multicasting, we have 
\begin{eqnarray}\label{eq:mellin}
	\mathcal{M}_{g(\gamma)}(s) &=& 1 + (s-1)\displaystyle\sum_{k_1+\ldots+k_M = K}\frac{\varphi\Gamma(1+\vartheta)}{\rho^{\vartheta}} \nonumber \\
	&& \hspace{-7mm} \times \ \Psi(\vartheta+1;\vartheta+s;K/\rho), \ \ \textrm{for} \ s < 1
\end{eqnarray}
	where $\Psi(a;b;z)$ is the confluent hypergeometric function of the second kind (also called Tricomi's confluent hypergeometric function \cite[Eq. 13.2.5]{Abramowitz1964} and denoted by $U(a,b,z)$), 
	\begin{equation*}
\varphi = \frac{\binom {K} {k_1,k_2,\ldots,k_M}}{\prod_{n=0}^{M-1}(n!)^{k_{n+1}}} \ \ \textrm{and} \ \ \vartheta = \textstyle \sum_{\ell = 0}^{M-1}\ell \cdot k_{\ell+1}.
	\end{equation*}
\end{theorem}


The above expression is quite complex and cumbersome to evaluate. The following lemma provides easily computable bounds using Alzer's inequality \cite{Alz97}. 

\begin{lemma}\label{lem1}
	The Mellin transform of $g(\gamma)$ can be bounded as 
	\begin{eqnarray}\label{eq:mellin_bound}
	1+(s-1)\mathcal{B}(s,b) \leq \mathcal{M}_{g(\gamma)}(s) \leq 1+(s-1)\mathcal{B}(s,1)
	\end{eqnarray}
where $b = [\Gamma(1+M)]^{-1/M}$ and
\begin{eqnarray*}
\mathcal{B}(s,\beta) & = & \textstyle \sum_{k=0}^{K}\sum_{j=0}^{kM}\binom{K}{k}\binom{kM}{j}(-1)^{k+j}e^{\frac{j\beta}{\rho}}\left(\frac{j\beta}{\rho}\right)^{1-s} \nonumber \\
&&\times \ \Gamma\left(s-1,\frac{j\beta}{\rho}\right).
\end{eqnarray*}
\end{lemma}

The above expressions and bounds provide a relatively accurate characterization of the service process and can easily be evaluated numerically. However, the quasi closed-form expressions are rather involved; they do not provide any insight on how the number of antennas and users affects the delay violation probability and its scaling. For that, we take a different approach and investigate the asymptotic behavior of the service process (and of its Mellin transform).

\begin{remark}
The above analysis allows us to directly obtain the effective capacity $\mathcal{R}(\theta)$ \cite{WuNegi03_EC}, i.e., the maximum constant arrival rate supported by the service process while satisfying statistical QoS requirements specified by the QoS exponent $\theta$, by noticing that 
\begin{equation}
\mathcal{R}(\theta) = -\frac{1}{\theta}\log\setM_{g(\gamma)}(1-\theta), \quad \theta>0.
\end{equation}
\end{remark}

\section{Delay Performance: Asymptotic Analysis} \label{sec:perf_exact}
In this section, we characterize the service process when $M$ is fixed, and $K$ is going to infinity. The first step is to find the asymptotic (limiting) distribution of the minimum SNR, which can be used to approximate its exact distribution.
\vspace{-0.5mm}

\subsection{Asymptotic Distribution}
We recall that the CDF of $X_{(1)}$ is
\begin{eqnarray}
L_K(x) = \P[X_{(1)} \leq x] = 1 - (1-F_{X}(x))^K.
\end{eqnarray}

From Fisher-Tippett-Gnedenko theorem \cite{David_EVT}, $F_{X}(x)$ belongs to the minimal domain of attraction of $L(x)$ if for at least one pair of sequences of real numbers ${c_K}$ and ${d_K > 0}$ it holds
\begin{eqnarray}
\lim_{K\to\infty}L_K(c_K+d_Kx) & = & \lim_{K\to\infty} 1 - (1-F_{X}(c_K+d_Kx))^K \nonumber \\
& = & L(x), \ \forall x.
\end{eqnarray}

Calculating the below necessary and sufficient condition 
\begin{eqnarray}
\lim_{\epsilon \to 0}\frac{F_X^{-1}(\epsilon) - F_X^{-1}(2\epsilon)}{F_X^{-1}(2\epsilon) - F_X^{-1}(4\epsilon)} = 2^{-\kappa},
\end{eqnarray}
we have that the shape parameter of the associated limit distribution $\kappa > 0$, which implies that $F_X(x)$ belongs to the Weibull minimal domain of attraction.
In other words, the CDF of $X_{(1)}$ converges to the scaled and translated Weibull CDF, denoted by $W(x)$, for sequences $\{c_K\}$ and $\{d_K > 0\}$, i.e.,
\begin{eqnarray*}
F_{X_{(1)}}(u) & = & W\left(\frac{u-c_K}{d_K}\right) \nonumber \\
& \to &  1 - \exp\left(-\left(\frac{u-c_K}{d_K}\right)^{\kappa}\right), \ u \geq c_K.
\end{eqnarray*}
The location constant $c_K$ is related to the lower end of the CDF $F_X(x)$ and is given as $c_K = v(F) = \inf\{x|F_X(x)>0\} = 0$, $\forall K$ since the chi-squared distribution is supported on $[0,\infty)$.  
\smallskip

The shape parameter $\kappa$ can be alternatively calculated as \cite{Leadbetter}
\begin{eqnarray}
\displaystyle \lim_{t\to\-\infty} \frac{F_X(v(F)-1/tx)}{F_X(v(F)-1/t)}=x^{-\kappa}
\end{eqnarray}
where evaluating the limit with $v(F) = 0$ gives $\kappa = M$.
\smallskip

The scale parameter is given by $d_K = F_X^{-1}(1/K) - v(F) = F_X^{-1}(1/K)$. Otherwise stated, we need to find $z$ such that $F_X(z) = 1/K$. Since $1/K$ approaches a very small value as $K \to \infty$, $z$ should be very small as well. So, approximating $F_X(x)$ with its Taylor expansion and keeping only the first term of the series, we have
\begin{eqnarray}
d_K = \frac{1}{M}\left[\frac{M!}{K}\right]^{1/M}.
\end{eqnarray}

Therefore, the limiting distribution of $X_{(1)}$ is 
\begin{eqnarray}
F_{X_{(1)}}(c_K + d_K x) = \P(X_{(1)} < d_K x) \stackrel{K \to \infty}{\longrightarrow} 1 - e^{-x^M}.
\end{eqnarray}
The support of the asymptotic distribution is 
\begin{eqnarray*}
S_L = \Big\{u\in[0,1]: \left(1-d_K(\log\frac{1}{\epsilon})^{\frac{1}{M}}\right) \leq u \nonumber \\ 
\leq \left(1-d_K(\log\frac{1}{1-\epsilon})^{\frac{1}{M}}\right)\Big\}
\end{eqnarray*}
where $\epsilon > 0$ is a very small number.

To quantify the accuracy of using the limit distribution for moderate number of users, we need to find a bound on the approximation/replacement error. We can show that $\P(X_{(1)} < d_K x) < 1 - e^{-x^M}$ and that the speed of convergence is faster than $\Theta(K^{-1/M})$. Using elementary results from \cite{Galambos} and after some algebraic manipulations, we have that
\begin{eqnarray*}
\left|\P(X_{(1)} < d_K x) - (1 - e^{-x^M}) \right| < e^{-x^M}x^{M+1}\left(\frac{M!}{K}\right)^{\frac{1}{M}}.
\end{eqnarray*}
 
Replacing the exact distribution by its asymptotic distribution, the Mellin transform of the service process for increasing $K$ is given by
\begin{eqnarray}\label{eq:mellin_as}
	\mathcal{M}_{g(\gamma)}^{\rm as}(s)  = 1+(s-1)\int_{0}^{\infty}(1+\rho x)^{s-2}e^{-(x/d_K)^M} \d x.
\end{eqnarray}

\subsection{Scaling results}
We present here results on the order growth of the service process when $M$ and/or $K$ is taken to infinity.
The easiest way is to derive an upper bound on the Mellin transform (using Jensen's inequality) and show that it is asymptotically tight.
\begin{theorem}
Let $\{X_k\}$ be positive random variables with finite mean and variance, and $\frac{|c_K|}{d_K} \to \infty$, then as $K\to\infty$
\begin{eqnarray}
f(\E[(X_{(1)}]) - \E[f(X_{(1)})] \to 0
\end{eqnarray}
where $f(x) = g(x)^{s-1}$.
\end{theorem}
The above convergence result implies that in order to calculate the Mellin transform of the service process, it is sufficient to evaluate the asymptotic mean of the minimum SNR. Note that the mean and the variance of Weibull distribution is given by $\E[W] = d_K\Gamma(1+1/M)$ and $\textrm{Var}[W] = d_K^2(\Gamma(1+2/M) - \Gamma^2(1+1/M))$, respectively.

\smallskip

\subsubsection{Finite $M$, Increasing $K$}
For MISO multicasting, as the number of users grows large, we have
	\begin{eqnarray}
	\lim_{K\to\infty}\setM_{g(\gamma)}^{\as}(s) \to \left(1+\rho\left(\frac{M!}{K}\right)^{\frac{1}{M}}\right)^{s-1} \approx O(K^{-\frac{1}{M}}).
	\end{eqnarray}

\subsubsection{Finite $K$, Increasing $M$}
For MISO multicasting, as the number of antennas grows large, we have
	\begin{eqnarray}
	\lim_{M\to\infty}\setM_{g(\gamma)}^{\as}(s) \to (1+P)^{s-1} \approx O(1).
	\end{eqnarray}

\subsubsection{Increasing $M$ and $K$}
We consider now the case where both the number of users and transmit antennas increase while maintaining a linear constant $\delta = K/M > 0$. For $\ell \in (0,1)$ and using Chebyshev's inequality and the fact that $\|\mathbf{h}_{k,i}\|^2/M$ has mean 1 and variance $1/2M$, we have
\begin{eqnarray}
\P(X_{(1)} \geq M\ell) \geq \left(1-\frac{1}{2M(1-\ell)^2}\right)^K \to e^{-\frac{\delta}{2(1-\ell)^2}}.  
\end{eqnarray} 

The Mellin transform can be lower bounded as follows
\begin{eqnarray}
\setM_{g(\gamma)}(s) & \geq & \P(X_{(1)} \geq M\ell) (1+P\ell)^{s-1} \\ 
& \rightarrow & e^{-\frac{\delta}{2(1-\ell)^2}}(1+P\ell)^{s-1} > 0.
\end{eqnarray}

Note that the service process is upper bounded by the multicast capacity (with perfect CSI), in which case the following upper bound on the Mellin transform can be found
\begin{eqnarray}
\setM_{g(\gamma)}(s) \leq  (1+P(1+\sqrt{\delta})^2)^{s-1} \approx O(1).
\end{eqnarray}

\section{Simulation Results} \label{sec:num}
In this section, we validate our delay performance analysis using simulations. The duration of a slot is set to 2 ms and the blocklength is $N=100$ symbols per slot. 

\begin{figure}[ht]
	\centering
	\includegraphics[width=0.9\columnwidth]{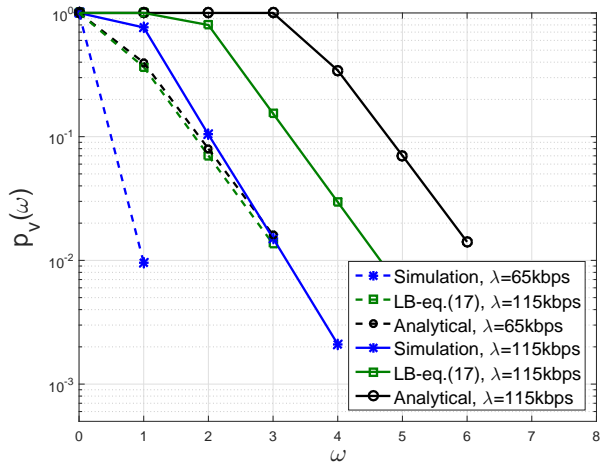}%
	\caption{Delay violation probability and associated bounds as a function of the target delay for different arrival rates, $M = 5$, $K=10$, and $P = 10$ dB.}
	\label{fig1}
\end{figure}

In Figure~\ref{fig1}, we compare the analytical expression on the delay violation probability and its lower bound with the simulated delay performance. We observe that the analytical expression curve follow quite well the trend of the simulated one, having a difference of about two slots. Furthermore, the proposed bound on $\setM_{g(\gamma)}$ given in (\ref{eq:mellin_bound}) has a smaller gap compared to the simulated performance. 

\begin{figure}[ht]
	\centering
	\includegraphics[width=0.9\columnwidth]{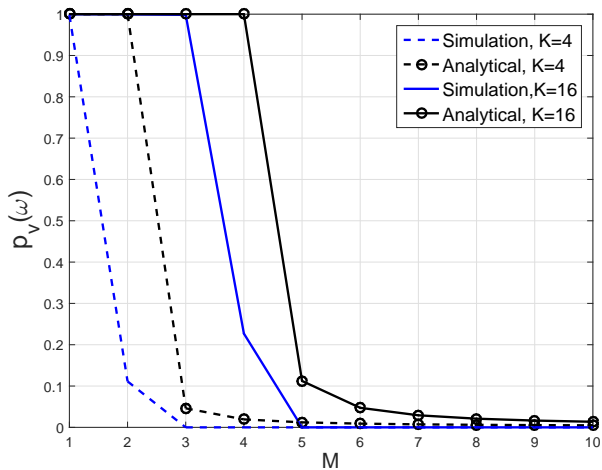}%
	\caption{Delay violation probability vs. number of antennas for $P = 10$ dB, arrival rate $\lambda = 100$ kbps, and $\omega = 3$ slots.}
		\label{fig2}
\end{figure}

In Figure~\ref{fig2}, we study the effect of the number of transmit antennas on the delay performance. Interestingly, for the scenario considered here, adding one antenna leads to a drastic drop of the delay violation probability.

\begin{figure}[ht]
	\centering
	\includegraphics[width=0.9\columnwidth]{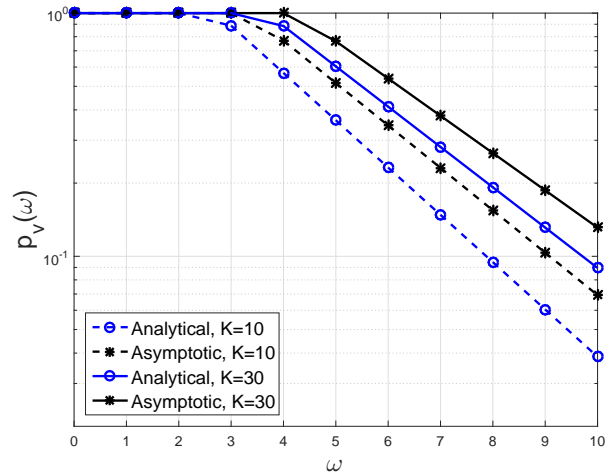}%
	\caption{Delay violation probability as a function of the target delay using the asymptotic distribution for $M = 10$, $P = 1$ dB, and $\lambda = 7.2$ kbps.}
	\label{fig3}
\end{figure}

Finally, in Figure~\ref{fig3}, we assess the effectiveness of our asymptotic analysis for charactering the delay violation probability. It can be seen that the asymptotic expression on $\setM_{g(\gamma)}$ provides satisfactory results even for moderate number of users. Moreover, the horizontal offset between the curve corresponding to the asymptotic delay violation probability and that of the non-asymptotic expression is of the order of one slot.  

\section{Conclusions} \label{sec:conc}
In this work, we investigated the queueing performance of physical layer MISO multicasting under statistical delay constraints. Using stochastic networks calculus, we derived a statistical characterization of the multicast service process and provided tight bounds on the delay violation probability. Furthermore, using extreme value theory, we characterized the service process for increasing number of users and provided scaling laws as the number of antennas and/or users is taken to infinity. Our analytical results indicate how the number of antennas, the number of users, and the transmit SNR may affect the delay violation probability in different system operating regimes.

%

\addcontentsline{toc}{chapter}{References}
\bibliographystyle{IEEEtran}
\bibliography{IEEEabrv,ref_Huawei,bib_snc,books_and_others}

\begin{thebibliography}{10}
\providecommand{\url}[1]{#1}
\csname url@samestyle\endcsname
\providecommand{\newblock}{\relax}
\providecommand{\bibinfo}[2]{#2}
\providecommand{\BIBentrySTDinterwordspacing}{\spaceskip=0pt\relax}
\providecommand{\BIBentryALTinterwordstretchfactor}{4}
\providecommand{\BIBentryALTinterwordspacing}{\spaceskip=\fontdimen2\font plus
\BIBentryALTinterwordstretchfactor\fontdimen3\font minus
  \fontdimen4\font\relax}
\providecommand{\BIBforeignlanguage}[2]{{%
\expandafter\ifx\csname l@#1\endcsname\relax
\typeout{** WARNING: IEEEtran.bst: No hyphenation pattern has been}%
\typeout{** loaded for the language `#1'. Using the pattern for}%
\typeout{** the default language instead.}%
\else
\language=\csname l@#1\endcsname
\fi
#2}}
\providecommand{\BIBdecl}{\relax}
\BIBdecl

\bibitem{eMBMS}
D.~Lecompte and F.~Gabin, ``Evolved multimedia broadcast/multicast service
  ({eMBMS}) in{ LTE}-advanced: overview and {Rel}-11 enhancements,'' vol.~50,
  no.~11, pp. 68--74, Nov. 2012.

\bibitem{DVB}
G.~Faria, J.~A. Henriksson, E.~Stare, and P.~Talmola, ``{DVB-H}: Digital
  broadcast services to handheld devices,'' \emph{Proceedings of the IEEE},
  vol.~94, no.~1, pp. 194--209, Jan. 2006.

\bibitem{Jindal}
N.~Jindal and Z.~Q. Luo, ``Capacity limits of multiple antenna multicast,'' in
  \emph{2006 IEEE Int. Symp. on Inform. Theory (ISIT)}, July 2006.

\bibitem{Park}
S.~Y. Park and D.~J. Love, ``Capacity limits of multiple antenna multicasting
  using antenna subset selection,'' \emph{{IEEE} Trans. Signal Process.},
  vol.~56, no.~6, pp. 2524--2534, June 2008.

\bibitem{SidiropoulosPMC}
N.~D. Sidiropoulos, T.~N. Davidson, and Z.-Q. Luo, ``Transmit beamforming for
  physical-layer multicasting,'' \emph{{IEEE} Trans. Signal Process.}, vol.~54,
  no.~6, pp. 2239--2251, June 2006.

\bibitem{KaripidisMC}
E.~Karipidis, N.~D. Sidiropoulos, and Z.~Q. Luo, ``Quality of service and
  max-min fair transmit beamforming to multiple cochannel multicast groups,''
  \emph{{IEEE} Trans. Signal Process.}, vol.~56, no.~3, pp. 1268--1279, Mar.
  2008.

\bibitem{CE_PLM}
S.~Zhang, R.~Zhang, and T.~J. Lim, ``{MISO} multicasting with constant envelope
  precoding,'' \emph{IEEE Wireless Communications Letters}, vol.~5, no.~6, pp.
  588--591, Dec. 2016.

\bibitem{Chang00}
C.-S. Chang, \emph{Performance Guarantees in Communication Networks}.\hskip 1em
  plus 0.5em minus 0.4em\relax London, UK: Springer-Verlag, 2000.

\bibitem{Multihop13_Infocom}
H.~Al-Zubaidy, J.~Liebeherr, and A.~Burchard, ``A (min, x) network calculus for
  multi-hop fading channels,'' in \emph{Proc. IEEE INFOCOM}, Apr. 2013, pp.
  1833--1841.

\bibitem{Fidler15_Guide}
M.~Fidler and A.~Rizk, ``A guide to the stochastic network calculus,''
  \emph{{IEEE} Commun. Surveys Tuts.}, vol.~17, no.~1, pp. 92--105, First
  quarter 2015.

\bibitem{Abramowitz1964}
M.~Abramowitz and I.~A. Stegun, \emph{Handbook of Mathematical Functions with
  Formulas, Graphs, and Mathematical Tables}.\hskip 1em plus 0.5em minus
  0.4em\relax New York: Dover, 1964.

\bibitem{Alz97}
H.~Alzer, ``On some inequalities for the incomplete {G}amma function,''
  \emph{AMS Mathematics of Computation}, vol.~66, no. 218, pp. 771--778, Apr.
  1997.

\bibitem{WuNegi03_EC}
D.~Wu and R.~Negi, ``Effective capacity: a wireless link model for support of
  quality of service,'' \emph{{IEEE} Trans. Wireless Commun.}, vol.~2, no.~4,
  pp. 630--643, July 2003.

\bibitem{David_EVT}
H.~A. David and H.~N. Nagaraja.\hskip 1em plus 0.5em minus 0.4em\relax Hoboken,
  NJ, USA: John Wiley \& Sons, Inc., 2005.

\bibitem{Leadbetter}
M.~R. Leadbetter, G.~Lindgren, and H.~Rootzen, \emph{Extremes and Related
  Properties of Random Sequences and Processes}.\hskip 1em plus 0.5em minus
  0.4em\relax New York, NY, USA: Springer, 1983.

\bibitem{Galambos}
J.~Galambos, \emph{The Asymptotic Theory of Extreme Order Statistics}.\hskip
  1em plus 0.5em minus 0.4em\relax Malabar, FL, USA: Krieger Publishing Co.,
  1987.

\end{thebibliography}

\end{document}